\documentstyle[12pt,psfig,aaspp4]{article}
\begin{document}
\title{Using Perturbative Least Action to Recover Cosmological 
Initial Conditions}

\author{David M. Goldberg and David N. Spergel}
\authoremail{(goldberg,dns)@astro.princeton.edu}
\affil{Princeton University Observatory, Princeton, NJ 08544-1001}

\begin{abstract}
We introduce a new method for generating initial conditions consistent
with highly nonlinear observations of density and velocity fields.
Using a variant of the Least Action method, called Perturbative Least
Action (PLA), we show that it is possible to generate several
different sets of initial conditions, each of which will satisfy a set
of highly nonlinear observational constraints at the present day.  We
then discuss a code written to test and apply this method and present
the results of several simulations.
\end{abstract}


\section{Introduction}

What initial density fluctuations gave rise to the present day
structure in the universe?  We have a number of reasons for asking
this question.  First, generating consistent initial conditions for
observations on small scale would give us a means of extracting the
small-scale primordial power spectrum, which can, for example, be used
to constrain the neutrino mass (Hu, Eisenstein \& Tegmark 1998).  In
addition, many dynamic systems of interest such as groups and clusters
can be tested for consistency with cosmological models by determining
what initial conditions could give rise to them within a given
scenario.  Finally, initial conditions which will evolve to satisfy
known constraints at the present day can be useful as input to
numerical simulations that study the evolution of galaxies and
clusters.

If the primordial fluctuations were Gaussian, then the power spectrum
on large scale ($\gtrsim 10$ Mpc) today fully characterizes the
statistical properties of the cosmological density field (e.g. Peebles
1980 \S 10 and references therein).  Any linear field can be uniquely
time-reversed to provide an initial density field at high redshift.  A
linear field is one which, when smoothed on sufficiently large scales,
gives a standard deviation of density perturbations, $\delta(({\bf
x},t_0)\equiv \rho({\bf x},t)/\overline{\rho}(t_0)-1$, of less than
unity, where $t$ is the physical time, $t_0$ is time at $z=0$, and
where ${\bf x}$ is a position in space in comoving coordinates.  When
density fluctuations are small, linear perturbation theory yields the
following simple relationship between an observed distribution at the
present and at high redshifts:
\begin{equation}
\delta({\bf x},t_i)=\frac{D(t_i)}{D(t_0)}\delta({\bf x},t_0)\ ,
\end{equation}
where the linear growth factor at present, $D(t_0)$ is normally set to
unity.

The problem of determining initial conditions for highly nonlinear
final density fields is a much more complex problem.  Due to mode-mode
coupling in the evolution equations (see e.g. Peebles 1980 \S 18),
time-reversal of a set of orbits becomes fundamentally ill-posed; that
is, many different sets of initial conditions can give rise to the
same or similar final density fields.  In an N-body simulation, this
can be thought of in terms of the trajectory of particles crossing one
another.  Since a region of large overdensity is populated by many
particles originating elsewhere, assigning particles uniquely to their
point of origin becomes impossible.

In this paper, we will develop a method for dealing with the problem
of time-reversing highly non-linear gravitational dynamic systems in a
realistic physical context.  The basic goal throughout will be as
follows: given some observed or target density field, $\delta({\bf
x},t_0)$ or a set of target constraints, such as density peaks or
voids in particular places, how does one go about generating one or
more sets of initial conditions, $\delta({\bf x},t_i)$, which, when
run through a gravity code, will yield the desired final conditions?

In order to answer this question, in \S 2 we will describe generic
methods for going from general constraints to constraints on
trajectories of individual particles.  We will further discuss some of
the methods that other researchers have used to try to satisfy those
constraints and generate initial conditions.  In \S 3, we discuss one
of the most promising methods, Least Action analysis, which gives a
single correct, but not necessarily physically well motivated, set of
initial conditions.  In \S 4, we will describe Perturbative Least
Action (PLA), which allows one to generate well motivated initial
conditions by perturbing random realizations of a known initial power
spectrum.  In addition, we will discuss a set of codes which have been
written in order to apply PLA to some test fields.  In \S 5, we will
show the results of two groups of toy problems used to test the PLA
code.  Finally, in \S 6, we will consider future applications.

\section{The Cosmological Constrained Boundary Problem}

\subsection{Initial Conditions: The Zel'dovich Approximation}

In the standard cosmological model, perturbations in the density field
at the present day arose out of a nearly uniform field at early times.
Rather than treat the density field as a continuum, it is convenient
to think of the matter in the universe as a distribution of particles
which may be binned and smoothed in order to yield a density field.
Of course, the positions and velocities of those particles may be
evolved using any of the standard N-body techniques (e.g. Hockney
\& Eastwood 1981, and references therein).  

At very early times, the particle field deviates little from a uniform
grid.  For convenience, we will use the notation, ${\bf q}_i$ to
denote the positions of particles on the grid, and ${\bf
d}_i(t_i)\equiv {\bf d}({\bf q}_i,t_i)$ to denote the displacement of each
particle from its gridpoint as well as a vector field of the same.
Thus, at $t=t_i$:
\begin{equation}
{\bf x}_i(t_i)={\bf q}_i+{\bf d}_i(t_i)\ .
\end{equation}
However, it may readily be shown that for small perturbations:
\begin{equation}
\delta({\bf x},t_i)\propto -\nabla \cdot {\bf d}({\bf q}_i,t_i) \ .
\end{equation}

But, as we pointed out earlier, small perturbations grow according to
a linear growth factor.  Thus:
\begin{equation}
{\bf d}_i(t)=\frac{D(t)}{D(t_i)}{\bf d}_i(t_i)
\end{equation}
as long as perturbations remain small.  This is the well known
Zel'dovich approximation (Zel'dovich 1970).  In addition, since it may
be shown that fields for which there is a curl in ${\bf d}({\bf
q}_i,t_i)$ produce decaying modes and our preferred model contains
only growing modes in the ``linear'' regime, we will always assume
that ${\bf d}({\bf q}_i,t_i)$ may be expressed as the gradient of a
scalar field.

\subsection{Final Conditions: Matching a Set of Constraints}

\label{sec:match}

After a particle field has evolved via gravitational collapse, we may
once again measure the corresponding density field, $\delta({\bf
x},t_0)$.  However, since our concern in this exercise is ensuring
that the density field satisfies some set of constraints, we will
now discuss how to generate a particle field which satisfies density
field constraints.  

In most reconstruction schemes, the goal is to find the ``true''
initial conditions for some randomly selected region of the universe
(e.g. Narayanan \& Croft 1999 and references therein).  Normally,
tests of these methods consist in running an N-body simulation and
attempting to match the initial conditions by examining the final
conditions.  For this, a particle field, $\tilde{\bf x}_i(t_0)$, is
laid down such that the smoothed density field of the particles yields
the target density field.  At this point, no consideration is given to
where a particle started out.  However, recall that at $a=0$ (where
$a$ is the cosmological expansion factor, and $a(t_0)\equiv 1$)
particle $i$ necessarily sits at its gridpoint, ${\bf q}_i$.  Since we
do not want particles traveling inordinately far, and since the
smoothed density density field will remain unchanged if we interchange
the indices of two particles, it is general practice that particles
will be interchanged until they have to move as far as possible.
Thus, we find the permutation matrix ${\bf M}_{ij}$ such that
\begin{equation}
\Delta^2[{\bf M}_{ij}\tilde{\bf x}_j(t_0),{\bf q}_i]\equiv 
\sum_{i}\left[{\bf q}_i- {\bf M}_{ij}\tilde{\bf x}_j(t_0) \right]^2
\end{equation}
is minimized.  This can be done, for example, using a simulated
annealing method (Press et al. 1992).

We finally define
\begin{equation}
{\bf x}_i(t_0)={\bf M}_{ij}\tilde{\bf x}_j(t_0)
\end{equation}
as the final particle positions.  

In addition to problems which constrain the entire density field, we
are also interested in scenarios in which particular regions are
constrained to have particular overdensities.  We discuss this
alternate set of constraints \S~\ref{sec:target}.

\subsection{Matching the Initial and Final Conditions}

At this point, we have found an initial and final position for each
particle.  However, {\it initial} in this sense refers to the
particle's position at $a=0$.  In a practical sense, we are interested
in the particle's position shortly thereafter.  And it is this
high-redshift position and velocity (coupled via the Zel'dovich
approximation) which we shall henceforth refer to as the ``initial
conditions'' of a particle field.

The Perturbative Least Action (PLA) has been developed in order to
solve this problem in a new and physical well-motivated way.
Narayanan \& Croft (1999) discuss other attempts recover initial
conditions, including linear theory, the Zel'dovich-Bernoulli Method
(Nusser \& Dekel 1992), Gaussianization (Weinberg 1992), and PIZA
(Croft \& Gazta\~naga 1997), which they show as most accurately
reproducing the initial conditions.  PIZA essentially sets the initial
offsets of the particles as:
\begin{equation}
{\bf d}_i(t_i)=D(t_i)\left[{\bf x}_i(t_0)-{\bf q_i}\right]\ ,
\end{equation}
with a corresponding velocity given by the Zel'dovich approximation.

\section{Ordinary Least Action}

Another reconstruction scheme which has received a great deal of
attention is the Least Action approach (Peebles 1980 \S 20, 1989,
1993, 1994, Shaya, Peebles \& Tully 1995).  It is from these examples
that we will make our foray into Perturbative Least Action, and
therefore we will discuss this method in some detail.

The idea of Least Action is that, given a set of boundary constraints,
such as the initial and final positions of each particle, for example,
or the initial position, and the final angle and radial velocity, one
can determine the set of orbits of particles by finding the
trajectories that extremize the action, defined as the time integral
of the Lagrangian.

In a cosmological context, the Lagrangian of a particle, $i$, can be
expressed as:
\begin{equation}
{\cal L}_i=m_i\frac{{\bf u}_i^2}{2}-m_i\Phi_i\ ,
\end{equation}
where ${\bf u}_i\equiv \partial (a{\bf x)}_i/\partial t$ and $\Phi_i$ is the total potential (background plus
perturbations) at the position of particle $i$.  We will henceforth
assume that the mass, $m_i$, is the same for all particles, and for
convenience, we will units in which $m_i=1$.  By subtracting out the
Lagrangian of a particle in a homogeneous universe (see, e.g. Peebles
1980 \S 7 for a derivation), the Lagrangian reduces to:
\begin{equation}
{\cal L}_i=\frac{a^2\dot{\bf x}_{i}^2}{2}-\phi_i\ ,
\end{equation}
where $\phi_i$ is the potential felt by particle $i$ due to the
density perturbations alone, and is equal to zero in an infinite,
smooth density field.

The cosmological Least Action variational principle  states that a set of particles, each traveling between
two known points, will each take the path which locally extremizes the
action, defined as:
\begin{equation}
S\equiv \sum_i \int_0^{t_0} dt \ {\cal L}_i= \sum_i \int dt \left( 
\frac{a^2\dot{\bf x}_{i}^2}{2}-\frac{\phi_i}{2} \right)\ ,
\label{eq:actdef}
\end{equation}
From now on, we will dispense with limits on the time integrals since
all of them are implicitly from $t=0$ to $t=t_0$.

Though any extremum (minimum, maximum, or inflection point) will yield
physically viable orbits, it is numerically most stable to find the
set of orbits which locally produces the least action, which is the
approach and terminology we will use henceforth.

In the case of discrete point sources:
\begin{equation}
\phi_i\equiv \sum_j \phi_{ji}=\sum_j \frac{Gm_im_j}{r_{ij}}
\end{equation}
Since $\phi_{ij}=\phi_{ji}$ (as all masses are identical), the total
binding energy of the system is expressed as $\sum_i \phi_i/2$.

In order to minimize the action, we express each particle trajectory
as a linear combination of a set of basis functions, and then minimize
the action with respect to these coefficients:
\begin{equation}
{\bf x}_i(t)={\bf q}_i+f_0(t)\left[{\bf x}_i(t_0)-{\bf q}_i\right]+
\sum_{n=1}^{n_{max}} {\bf C}_{i,n}f_{n}(t)
\label{eq:lapath}
\end{equation}
where we have defined $f_{n=0,n_{max}}(0)=0$, $f_0(t_0)=1$,
$f_{n=1,n_{max}}(t_0)=0$, and $a^2\dot{f}_{n=0,n_{max}}(t)\rightarrow
0$ as $a\rightarrow 0$.  A good solution for the zeroth basis function
gives $f_0(t)=D(t)$.  Using these constraints, the Zel'dovich
approximation is necessarily satisfied for each basis function, and
hence, for each particle trajectory at early times.  

The Least Action principle demands that given a physical set of
orbits, all derivatives of $S$ with respect to ${\bf C}_{i,n}$
will vanish.  Thus:
\begin{equation}
\nabla_{{\bf C}_{i,n}}S=\int dt \left[\dot f_n(t) a^2
\dot {\bf x}_i-f_n(t)\nabla \phi_i\right]=0 \ ,
\label{eq:OLA}
\end{equation}
where here and throughout, unlabeled gradients are assumed to be with
respect to the comoving coordinate system.  By using the constraints
listed above and doing an integration by parts, this is algebraically
equivalent to:
\begin{equation}
\nabla_{{\bf C}_{i,n}}S=\int dt f_n(t) 
\left[-\frac{\partial}{\partial t}(a^2 {\bf \dot{x}})+\nabla
\phi_i\right]=0 \ .
\end{equation}
However, everything inside the parentheses on the right side of the
equation necessarily equals zero (as must its time integral) if the
equations of motion are satisfied, as it is merely Newton's second law
written in comoving coordinates.  By using the form of the trajectory
in equation~(\ref{eq:lapath}), we find a set of orbits which
necessarily satisfies both the equations of motion and the
constraints, and will converge quickly.

Since the evolution equations are implicitly dependent upon the
underlying cosmology, examination of the velocities of galaxies can
potentially give limits on cosmological parameters.  This approach has
been applied, for example, to galaxies within 3000 km/s (Shaya,
Peebles and Tully 1995; Dunn \& Laflamme 1995), yielding a value of
$\Omega_m\simeq 0.2$.  Branchini \& Carlberg (1995), on the other
hand, argue that Least Action analysis dramatically underestimates
$\Omega_m$, and that Local Group dynamics could yield a value as high
as $\Omega_m=1$.

In general, the ordinary Least Action approaches use direct
particle-particle summation to calculate the forces on particles.  We
use a particle mesh (PM) Poisson solver to compute forces, which
greatly speeds up computation.  A similar approach to ordinary Least
Action was used by Nusser and Branchini (1999) who used a tree code
scheme to compute particle forces.

While ordinary Least Action analyses provide physically correct orbits
for particles, the initial conditions found need not have any
resemblance to a field drawn from an {\it a priori} known power
spectrum.  Rather, particle trajectories have traditionally been
generated which evolve a field from a completely uniform one to one
satisfying the constraints using the least total distance for each
particle, as is the case with the PIZA algorithm.  In addition, the
actual path of the particles given by direct application of Least
Action (as well as linear perturbation theory or PIZA) essentially
gives a first infall solution, rather than allowing for the
possibility of orbit crossings.  Moreover, nothing in the generation
of initial conditions demands that the initial fields be curl-free;
thus, decaying modes can develop.

\section{Method: Perturbative Least Action}
\label{sec:PLA}

In order to alleviate these problems inherent in ordinary Least
Action, we now develop a method to generate an ensemble of initial
conditions, each as consistent as the constraints will allow with a
specified primordial power spectrum.  In this section we will first
develop the equations governing PLA.  We will then discuss the
background cosmology and numerical methods in our code.  Next, we will
discuss the various types of target density fields to be used in our
simulations.  We will then discuss how basis functions are generated.
Finally, we will explain how the perturbed action is minimized using
the PLA code.

\subsection{General Equations}

First, let us suppose that we have run an N-body code on a randomly
generated some set of initial conditions with known power spectrum.
The path of each particle, $ \{ {\bf x}^{(0)}_i(t) \} $, is known to
satisfy the cosmological equations of motion.  Let us, by perturbing
around the final ``unperturbed'' density field, find a set of ${\bf
x}_i(t_0)$ which produce a density field satisfying our constraints on
the system.  The full path of each particle can be expressed as a
perturbation around ${\bf x}_i^{(0)}(t)$ using the basis functions
introduced above.  Thus, we may say:
\begin{equation}
{\bf x}_i(t)={\bf x}_i^{(0)}(t)+f_0(t){\bf x}^{(1)}_i(t_0)+
\sum_{n=1}^{n_{max}} {\bf C}_{i,n}f_n(t)
\end{equation}
where we have applied the same constraints on $f_n(t)$ as discussed
above, and ${\bf x}^{(1)}_i(t)$ is the perturbation orbit such that
${\bf x}_i^{(1)}\equiv{\bf x}_i-{\bf x}_i^{(0)}$.   Comparison of this
equation with equation~(\ref{eq:lapath}) illustrates that Least Action
and PLA are quite similar, but perturb around different guesses for
the particle trajectory.

Since for highly nonlinear systems, there may be many minima of the
action which produce the correct final conditions, by perturbing away
from a known field which is consistent with a given power spectrum, we
are able to keep each realization as physically relevant as possible,
and find the ``closest'' local minimum in parameter space.

We may now rewrite the action (equation~\ref{eq:actdef}) as,
\begin{eqnarray}
S&=&\sum_{i}\int_{0}^{t_{0}} dt \left(
\frac{a^2 \dot{\bf x}^{(0)2}_{i}}{2}-\frac{\phi^{(0)}_{i}}{2} 
\right)+
\sum_{i}\int_{0}^{t_{0}} dt \left( 
a^2\dot{\bf x}_{i}^{(0)}\cdot\dot{\bf x}_{i}^{(1)}+
\frac{a^2 \dot{\bf x}^{(1)2}_{i}}{2}-\frac{\phi_{i}}{2}+
\frac{\phi^{(0)}_{i}}{2} 
\right)\nonumber \\
&=&S^{(0)}+\sum_{i}\int_{0}^{t_{0}} dt \left( 
a^2\dot{\bf x}_{i}^{(0)}\cdot\dot{\bf x}_{i}^{(1)}+
\frac{a^2 \dot{\bf x}^{(1)2}_{i}}{2}-\frac{\phi_{i}}{2}+
\frac{\phi^{(0)}_{i}}{2} 
\right)\ ,
\label{eq:PA}
\end{eqnarray}
where $\phi^{(0)}_i$ is the potential on particle, $i$, in the
unperturbed, (${\bf x^{(0)}}$), potential field.

The gradient of the action with respect to the coefficients is:
\begin{equation}
\nabla_{{\bf C}_{i,n}}S=\int dt \left[
\dot{f}_{n}(t)a^{2}\left(\dot{\bf x}^{(0)}+\dot{\bf x}^{(1)}\right)-
f_{n}\nabla\phi_{i}
\right]
\end{equation}
However, by definition, 
\begin{equation}
\int dt \left[ \dot f_{n}(t)a^2 \dot {\bf x}^{(0)}_i-f_n(t)\nabla \phi^{(0)}_i\right]=0 \
,
\end{equation}
so,
\begin{equation}
\nabla_{{\bf C}_{i,n}}S=\int dt \left[
\dot{f}_{n}(t)a^{2}\dot{\bf x}^{(1)}+f_{n}\left(
\nabla\phi^{(0)}_i-\nabla\phi_i
\right)
\right]
\label{eq:PLA}
\end{equation}

It is also often useful to calculate the Hessian Matrix of second
derivatives in order to minimize the action:
\begin{equation}
{\cal H}_{i\alpha\beta,nm}\equiv
\frac{\partial^2 S}{\partial C_{in}^{\alpha}\partial C_{jm}^{\beta}}=
\int dt \left[a^2 \dot{f}_n\dot{f}_m\delta_{ij}\delta_{\alpha\beta}-
f_n f_m\frac{\partial^2 \phi_i}{\partial x_{i}^\alpha \partial
x_{j}^{\beta}}
\right]\ ,
\label{eq:dc2ds2}
\end{equation}
where $\alpha$ and $\beta$ are direction indices.  We will discuss
application of the Hessian matrix for minimization in \S~\ref{sec:minimize}.

\subsection{The PLA Procedure}
\label{sec:outline}

In this section, we describe how PLA is actually applied in practice.
We begin the process by running an N-body simulation with a random
seed.  These trajectories will be referred to as ${\bf x}_i^{(0)}(t)$.
In our simulations, we use a Particle Mesh (PM) code (Hockney \&
Eastwood 1981).  Though the power spectrum and cosmology of the
unperturbed simulation is held to be constant in the following
simulations, in a forthcoming paper, we will show how PLA may be used
to discriminate between different cosmological models.

\subsubsection{The Target Density Field}
\label{sec:target}

After we run the unperturbed simulation, we must next figure out what
perturbations need to be applied to the particle paths at $t=t_0$.
This may be done in any way, may satisfy any sort of constraint, and
the details of determining the final positions of the perturbed orbits
are not crucial to the PLA method itself.  However, since some
perturbations will be easier to satisfy than others, we will briefly
discuss the method used in our code to perturb the particle
trajectories.  We will thus discuss means of generating a target
density field, $\delta({\bf x},t_0)$, from the smoothed density field
of the unperturbed simulation, $\delta^{(0)}({\bf x},t_0)$.

In \S~\ref{sec:match}, we discussed the traditional method of
generating a target field: running a numerical simulation with a
different random seed than our unperturbed field.  However, the PLA
approach was originally formulated with the intent of providing
initial conditions to highly nonlinear individual constraints, such as
rich clusters appearing in particular regions, or large voids
appearing elsewhere.  We will now discuss how to generate a
``realistic'' target density field which constrains the mean final
overdensity in particular regions, which resembles the unperturbed
simulation as closely as possible.

To this end, we use a very similar approach to the ``Constrained
Initial Conditions'' method employed by Hoffman \& Ribak (1991, 1992),
and pioneered by Bertschinger (1987).  While Constrained Initial
Conditions are actually meant to provide initial conditions on large
scales (linear at the present day), as a side effect, it can be used
to map one (even highly nonlinear) density field onto another while
satisfying a set of given constraints and at the same time preserving
the autocorrelation of the density field.

While objections might certainly be made that the Constrained Initial
Conditions method makes assumptions about the statistical properties
of the density field to be generated (such as that it is Gaussian
random) which may not hold in the nonlinear regime, we re-iterate that
we have chosen this method for generating a final target field because
it is convenient, but not intrinsic to the PLA method, itself.  

In order to create a ``target'' density field (the field which will
both satisfy the given constraints and has the same large scale
distribution as the unperturbed field), we begin by computing a
gridded density field, using a Cloud-In-Cell (CIC) interpolation
scheme (Hockney \& Eastwood 1981).  Next, we calculate the
autocorrelation function $\xi^{(0)}(r)$ of the unperturbed field,
$\delta^{(0)}({\bf x},t_0)$.

We define our constraints such that within some region, ${\cal R}_n$
(defined as a normalized tophat function), we want to have some given
mean, $\overline\delta({\cal R}_n,t_0)=c_n$.  Following the
prescription of Hoffman \& Ribak, we next compute the correlation of
each constraint with every point on the density field:
\begin{equation}
\xi^{(0)}_n({\bf r})=\int d^3{\bf r}' \xi^{(0)}(|{\bf r}-{\bf r}'|)
\delta^D({\bf r}'-{\cal R}_n)
\end{equation}

In addition, we need to compute the correlation between each
constraint:
\begin{equation}
\xi^{(0)}_{nm}=\xi^{(0)}_{mn}=\int d^3{\bf r} \xi^{(0)}_n({\bf r}) \delta^D({\bf
r}'-{\cal R}_m)
\end{equation}

We then constrain the density field by applying the relation:
\begin{equation}
\tilde\delta({\bf x},t_0)=\delta^{(0)}({\bf x},t_0)+\xi^{(0)}_n({\bf
r})\xi^{-1}_{nm}(c_m-c^{(0)}_m)\ .
\end{equation}

At this point, we choose to regularize our solution.  We have found it
practical to demand that the initial power spectrum of the field
remain unchanged on perturbation.  We thus apply a correction to the
initial perturbed field such that
\begin{equation}
\delta({\bf k},t_0)= \tilde\delta({\bf k},t_0) 
\sqrt{\frac{P^{(0)}(k)}{\tilde{P}(k)}} \ .
\label{eq:norm_power}
\end{equation}
This necessarily insures that the final density fields have the same
power spectrum, and generally only makes a correction at large scales
where linear theory would be expected to hold.  Others may choose to
add different forms of regularization to their simulations.  The
choice is not intrinsic to PLA, itself.

We may also compute a target density field by simply taking the
observed density field from another simulation, as we will do in the
second set of simulations.  

Once the target density field has been computed and particle positions
are determined which satisfy this field, it remains only to calculate
the permutation matrix, ${\bf M}_{ij}$ in order to have a set of final
constraints.  We may use the basic procedure discussed in
\S~\ref{sec:match}, with one small adjustment.  Rather than finding a
permutation matrix that minimizes offsets from the uniform grid, in
PLA we want to minimize $\Delta^2[{\bf M}_{ij}\tilde{\bf
x}_j(t_0),{\bf x}^{(0)}_i(t_0)$].  Recall
\begin{equation}
{\bf x}_i(t_0)={\bf M}_{ij}\tilde{\bf x}_j(t_0)\ ,
\end{equation}
and
\begin{equation}
{\bf x}^{(1)}_i(t_0)={\bf x}_i(t_0)-{\bf x}^{(0)}_i(t_0)\ .
\end{equation}
We have thus generated the sought-after perturbations for the final
particle field.

\subsubsection{Basis Functions}

Once we have generated unperturbed and perturbed positions, we next
run the PLA code which will determine initial conditions which will
give rise to the perturbed field.

The first step in this code is determination of appropriate basis
functions.  A natural choice is that all basis functions should be
polynomials with base $D(t)$, since we know that lower order
perturbation theory will yield solutions of this form.  Earlier, we
said that $f_0=D(t)$.  Likewise, we set:
$f_1=D(t)\left[1-D(t)\right]$.

For higher order basis functions, we examine the unperturbed
trajectories, and successively create best fit basis function
(polynomials which best fit the residuals of the unperturbed
trajectories) of the form:
\begin{equation}
f_n=\sum_{m=n}^{m_{max}}b_{nm} D(t)^m\left[D_0-D(t)\right]\ ,
\end{equation}
where the kernel polynomials are the same form used by Giavalisco et
al. (1993).

Given the form of the coefficients, only $f_0$ and $f_1$ grow linearly
at early times, and hence, the corresponding coefficients are those
which are used to generate the initial conditions.

\subsubsection{Determining the Coefficients}

\label{sec:minimize}

We are now prepared to compute the coefficients which minimize the
action.  Because of the form of the interpolation scheme for the
potential in PM codes, computation of the actual potential, itself,
rather than its spatial derivatives is not well defined.  Hence, when
we minimize the action we actually want to find coefficients such that
the derivatives of the action (equation~\ref{eq:PLA}) vanish.

In order to do this, we assume that each particle and direction are
approximately independent of one another, and that the strongest
correlations will be between different coefficients of the same
particle.  Our minimization scheme is similar to the
Levenberg-Marquardt Method (Press et al. 1992, \S 15.5), and uses the
inverse of the Hessian matrix to compute the steps in coefficients:
\begin{equation}
\Delta C^\alpha_{i,n}=-\sum_{m=1}^{n_{coeff}}\frac{\partial S}{\partial
C^\alpha_{i,m}}{\cal H}^{-1}_{i\alpha\alpha,mn} \ .
\label{eq:step_est}
\end{equation}

In addition to minimizing the action, we wish to put a constraint on
our trajectories that they only have a growing mode at early times.
Consistent with that is our assumption that our initial velocity field
be curl free.  In order to insure this, we must confine ${\bf
C}_{i,1}$ to a submanifold such that
\begin{equation}
\nabla_{\bf q} \times {\bf C}_{i,1}=\nabla_{\bf q} \times {\bf \Delta
C}_{i,1}=0\ .
\end{equation}
Recall that of the basis functions, only the first grows linearly at
early times, and hence, at high redshift, the velocity field will be
given by the first coefficients.  Further recall that at early times,
the positions of particle, $i$, is approximately given by ${\bf q}_i$,
allowing us to take spatial derivatives of the velocity field.

In order to make $\Delta {\bf C}_{i,1}$ curl-free, we take the Fourier
transform of the initial estimate of the step
(equation~\ref{eq:step_est}), and find a new Field such that
$\Delta {\bf C}_{1}({\bf k})\propto {\bf k}$ everywhere, and the
difference squared of the old and new field are minimized.  

We then iterate until convergence is reached.  At this point, the
initial positions and velocities of the perturbed field are computed,
and the initial density field is corrected to yield an identical power
spectrum to that of the unperturbed field using
equation~(\ref{eq:norm_power}). This new field may be plugged into an
N-body code and will approximately yield ${\bf x}(t_0)$.

Of course, if satisfactory convergence is not reached after one try,
the results of the final N-body simulation may be used as the {\em
new} unperturbed simulation, and the process may be repeated as needed. 

\section{The Simulations}

In order to test the concept of the PLA method, we have run two groups
of simulations.  A discussion of the numerical method is presented in
the previous section.

\subsection{Test 1: Matching a Random Field}

The first test of the PLA code is a natural choice for any code which
tries to determine initial conditions; namely, given a density field
from an N-body simulation, how well can the method compute the initial
conditions.  This is the test constructed by Narayanan \& Croft
(1999), in which they demonstrate the superiority of PIZA to other
reconstruction schemes.

In our test, we run a randomly realized simulation with standard CDM
cosmology ($\Omega_0=1.0$, $\Lambda=0.0$) on a grid with $64^3$
particles, $128^3$ gridcells, a gridlength of $100 h^{-1}$ Mpc, and
with $\sigma_8=1$, and compute the final density field.  We then run
another simulation with the same cosmology, which will be used for
nothing more than generating an initial power spectrum.  Objections
might be raised that we have no {\it a priori} knowledge of the
cosmology, and hence, have no right to do this.  However, we will show
in a forthcoming paper how one may use PLA to generate a maximum
likelihood estimate of cosmological parameters.  For now, we will
assume such parameters are known (as they are assumed to be in
Naryanan and Croft 1999).  

From the target density field, we then compute a final particle field,
${\bf x}_i(t_0)$, which both satisfies the density field, and
minimizes $\Delta^2[{\bf q}_i,{\bf x}_i(t_0)]$.  Note that we are
perturbing away from a uniform field, rather than from the positions
of a randomly generated simulation.  From here, however, we apply the
method exactly as described in \S~\ref{sec:PLA}.

With this test we hope to examine two things: 1) How well are the
actual initial conditions from the target field reproduced?  2) After
running the generated initial conditions through the PM code, how well
does the final density field match the final target density field?  

Recall that as part of the method, we are constraining the initial
power spectrum to match the ``known'' primordial power spectrum.  In
addition, our constraints guarantee that the field is necessarily curl
free, and that the initial displacements satisfy the Zel'dovich
approximation.  Beyond that, a good statistical test of difference is
given by Narayanan \& Croft (1999), who define a difference of complex
amplitudes between field ``1'' and ``2'' as: 
\begin{equation}
D(k,t)=\frac{\sum [\delta_1({\bf k})-\delta_2({\bf k})]^2}
{\sum [\delta_1({\bf k})^2+\delta_2({\bf k})]^2}\ ,
\end{equation}
where $\delta_1({\bf k})$ and $\delta_2({\bf k})$ are the Fourier
components of density field ``1'' and ``2'' respectively.  

\begin{figure}
\centerline{\psfig{figure=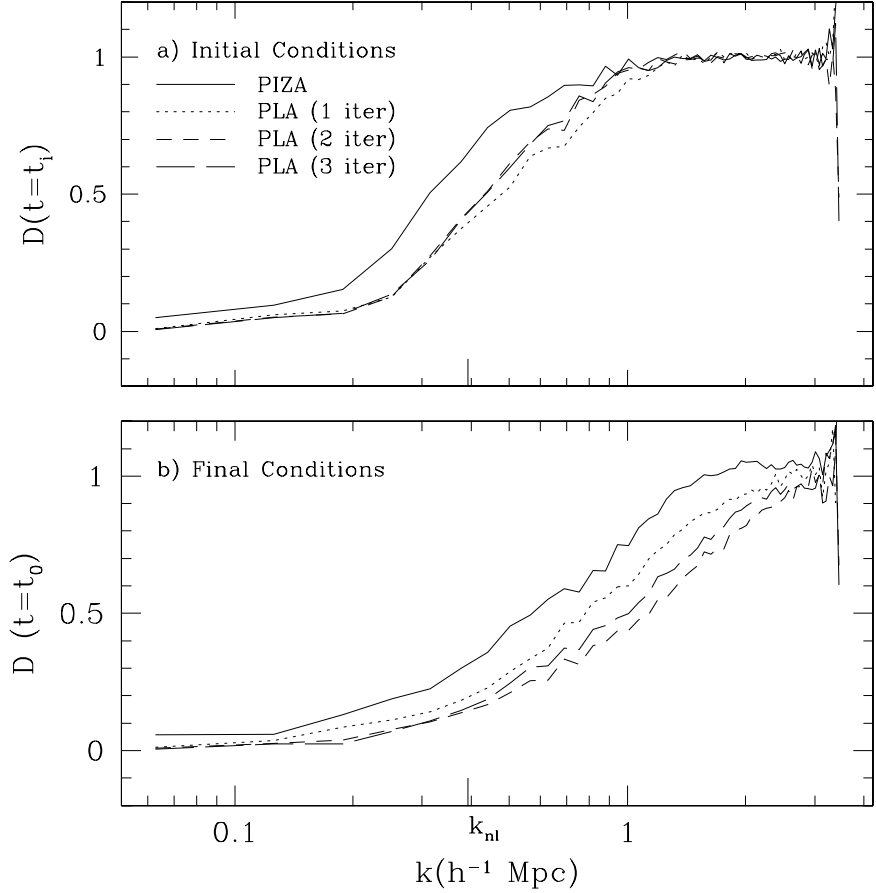,angle=0,height=6in}}
\caption{a) Comparison of the initial density fields generated using
PIZA (solid line), and various iterations of PLA (dotted, short
dashed, and long dashed lines for 1, 2, and 3 iterations,
respectively), to the actual initial conditions used to generate a
random target field.  The density fields are Fourier decomposed, and
the modes are compared according to the difference coefficient,
$D(k)=\sum[ \delta_1({\bf k})-\delta_2({\bf k})]/\sum[\delta_1({\bf
k})^2+\delta_2({\bf k})^2]$.  The nonlinear scale ($k=2\pi/16h^{-1}$
Mpc), is denoted by $k_{nl}$.  b) A similar comparison, but for the
evolved density fields.}
\label{fg:compare_sim}
\end{figure}

In Figure~\ref{fg:compare_sim}, we examine this statistic as a
function of k for both the initial and final condition fields.  We
look at the difference between the target field and PIZA, and various
numbers of iterations of PLA.  This is intended to be almost identical
to Figure 7 in Narayanan \& Croft (1999), and illustrates that even
with only one iteration, PLA fits both the initial and final target
density field better than PIZA, which was shown to be the previous
leading contender for initial condition reconstruction.

With further iterations, the initial field stabilizes at a somewhat
worse fit than the first iteration.  This is not two surprising since
the differences in fit occur exclusively on the nonlinear scale.
Since a unique fit cannot to the final field cannot be found, and
since the final field is ultimately what PLA tries to match, some
phase mixing may cause a slightly worse fit in the initial field.

The final field, however, improves on one additional iteration, and
then slightly worsens on a third.  Thereafter, we've found the result
more or less stabilizes.  We have found, however, that we get a better
fit on {\it all} physical scales for both the initial and final fields
by running our simulations in a box in which the resolution scale is
somewhat larger.

\subsection{Test 2: Constraining Clusters}

In our second set of simulations, we have tried to form three rich
clusters in specified positions within a known field. We have run three
randomly realized simulations with different random seeds.  Each is
run with a standard CDM cosmology ($\Omega_0=1.0$, $\Lambda=0.0$) on a
grid with $64^3$ particles, $128^3$ gridcells, a gridlength of $100
h^{-1}$ Mpc, and with $\sigma_8=1$.  We then specify three positions,
and demand that within a radius of $1.5 h^{-1}$ Mpc of those
positions, the mean overdensity become $\overline{\delta}=200$, giving
us rich Abell clusters.

The general method for doing this is described in great detail in \S
\ref{sec:PLA}.  In this test, we used 4 basis functions for each set
of simulations, and typically ran four cycles (minimizing of the
action, and re-running the new initial conditions in the PM code)
until satisfactory convergence was reached.

As illustrated in Table~\ref{tab:fit1} and \ref{tab:fit2}, the results
of this exercise are quite successful.  In the figures, the integrated
average density within the constraints regions is shown for the 3
unperturbed and perturbed realizations, respectively.  Recall that
both the perturbed and unperturbed realizations are the output of the
running the perturbed and unperturbed initial conditions through a PM
code.

\begin{table}
\begin{tabular}{|c|c|c|c|c|}
\hline
\multicolumn{5}{|c|}{Value of Constraints in Unperturbed Simulations} \\ \hline
Cluster & $\overline{\delta}$ (model) & $\overline{\delta}$ (sim 1)
& $\overline{\delta}$ (sim 2)
& $\overline{\delta}$ (sim 3)  \\ \hline\hline 
1 & 200 & -0.41 & -0.20 & -0.83 \\
2 & 200 & -0.83 & -0.99 & -0.83 \\
3 & 200 & -0.98 &  0.06 & -0.97 \\
\hline
\end{tabular}
\caption{A summary of how well the random realizations satisfy the
constraints to be imposed.  The constraints consist of a region of
radius $1.5h^{-1}$ Mpc, with a mean overdensity of
$\overline{\delta}=200$.  If anything, the regions selected are voids
in the unperturbed realizations.}
\label{tab:fit1}
\end{table}

\begin{table}
\begin{tabular}{|c|c|c|c|c|}
\hline
\multicolumn{5}{|c|}{Value of Constraints in Perturbed Simulations} \\ \hline
Cluster & $\overline{\delta}$ (model) & $\overline{\delta}$ (sim 1)
& $\overline{\delta}$ (sim 2)
& $\overline{\delta}$ (sim 3)  \\ \hline\hline 
1 & 200 & 201 & 189 & 141 \\
2 & 200 & 155 & 176 & 215 \\
3 & 200 & 206 & 226 & 191 \\
\hline
\end{tabular}
\caption{A summary of how well the evolved perturbed fields satisfy
the  imposed constraints at $z=0$.  Note that the largest error in mass is
$\sim 30\%$, and that a typical error is about 10\%.}
\label{tab:fit2}
\end{table}

\begin{figure}
\centerline{\psfig{figure=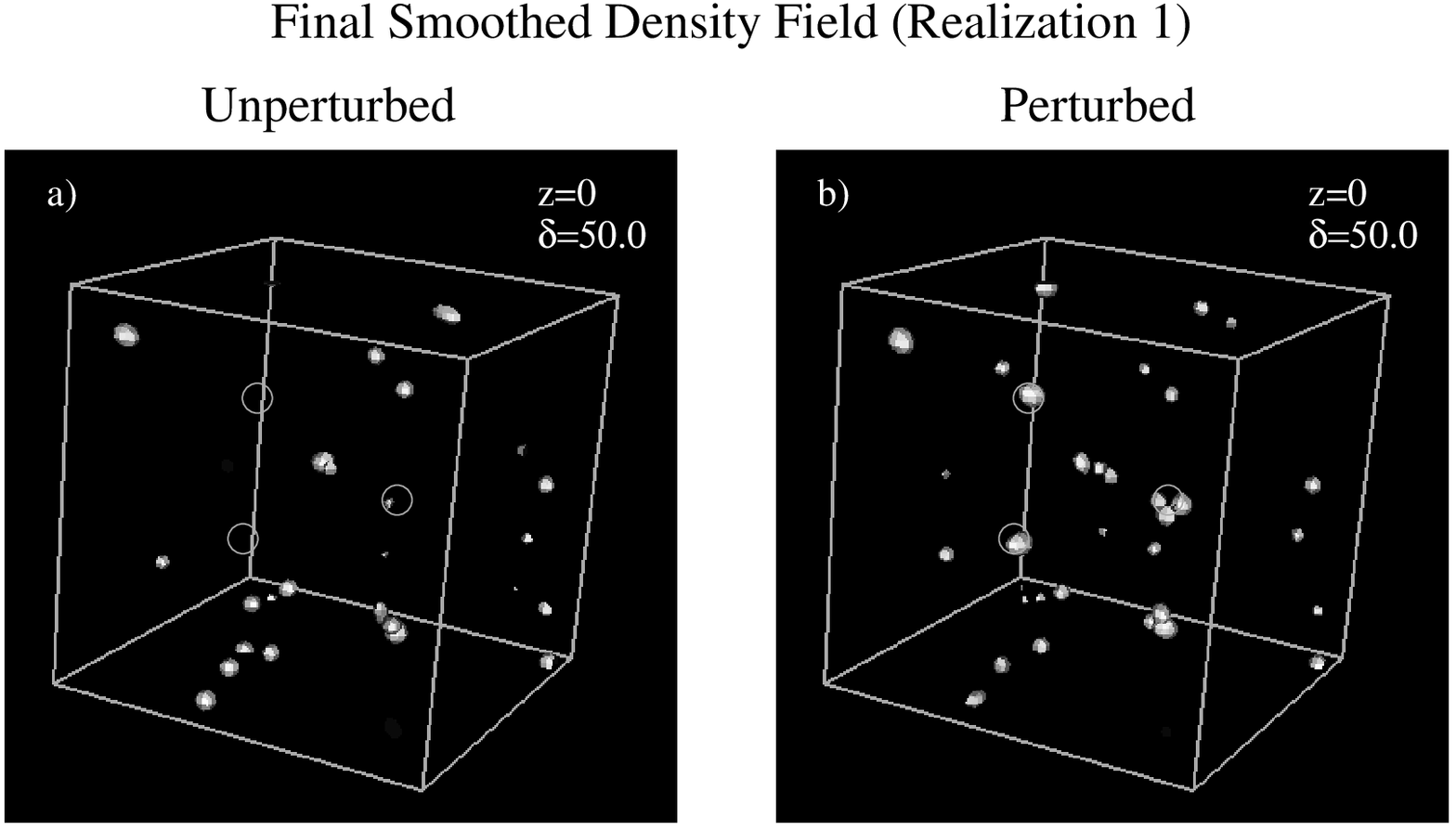,angle=-90,height=4in}}
\caption{Realization 1.  a) A plot the smoothed ($r=3 h^{-1}$Mpc)
density field, showing contours of $\delta=50$ for the first
unperturbed realization. b) A similar plot, showing the evolved
density field of the perturbed simulations.  The circles in each
indicate the location of the target clusters.  }
\label{fg:sim1}
\end{figure}

In addition to a strict evaluation of how well the constraints are
satisfied, a visual inspection of the final density fields may also be
instructive.  In Figures~\ref{fg:sim1}-\ref{fg:sim3} we show the
density fields of the a) unperturbed and b) perturbed evolved density
fields.  As a reminder, both are the result of actually running an
initial field through the PM code.  The plots show contours of regions
with overdensities of greater than 50 smoothed on a scale of $1.5
h^{-1}$ Mpc.  Note that though a number of the ``clusters'' remain
virtually unchanged through the perturbation, our rich target clusters
appear precisely on target in all three realizations.

\begin{figure}
\centerline{\psfig{figure=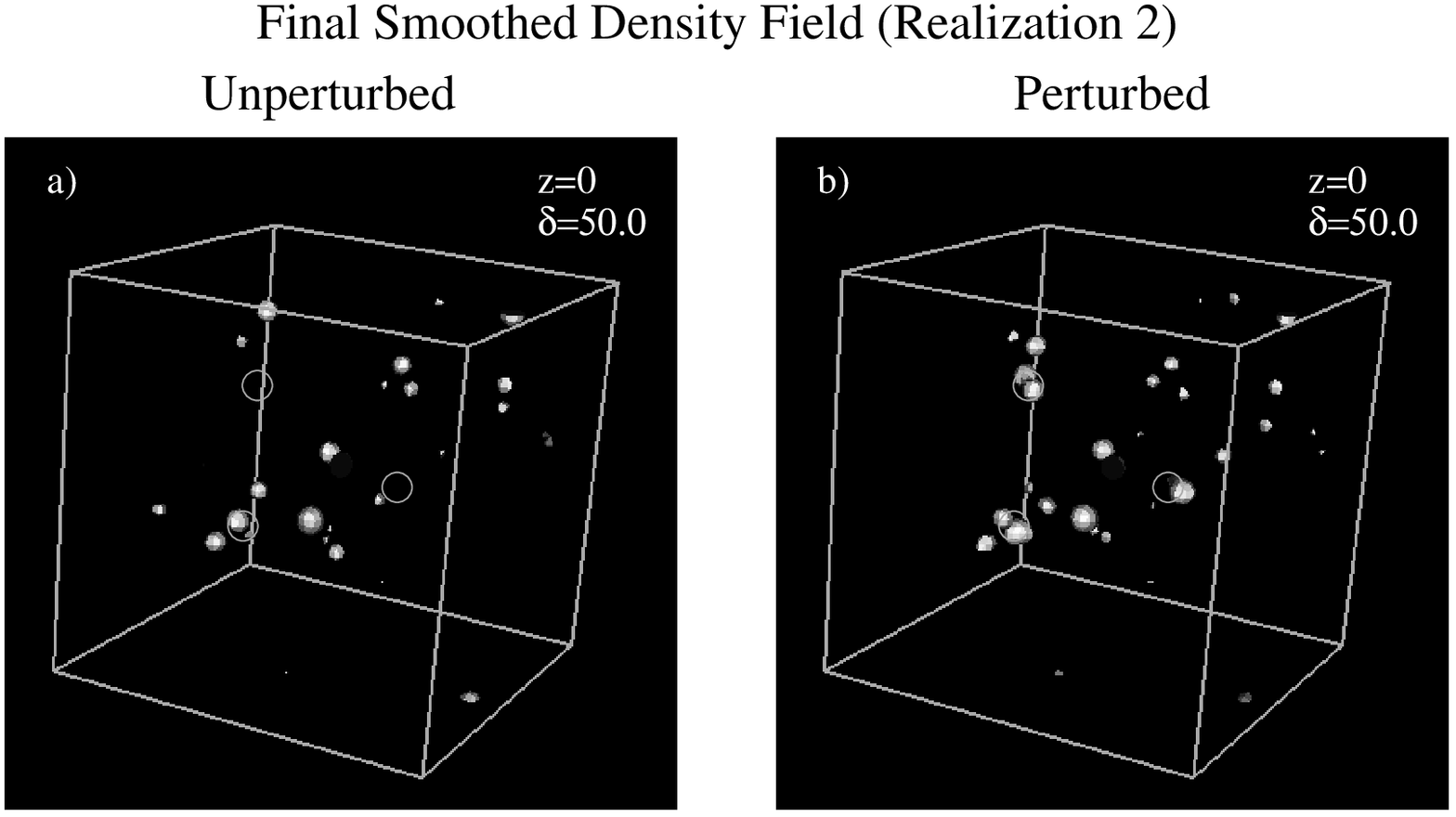,angle=-90,height=4in}}
\caption{Realization 2. As in Fig.~\ref{fg:sim1}, but for the second realization.}
\end{figure}

\begin{figure}
\centerline{\psfig{figure=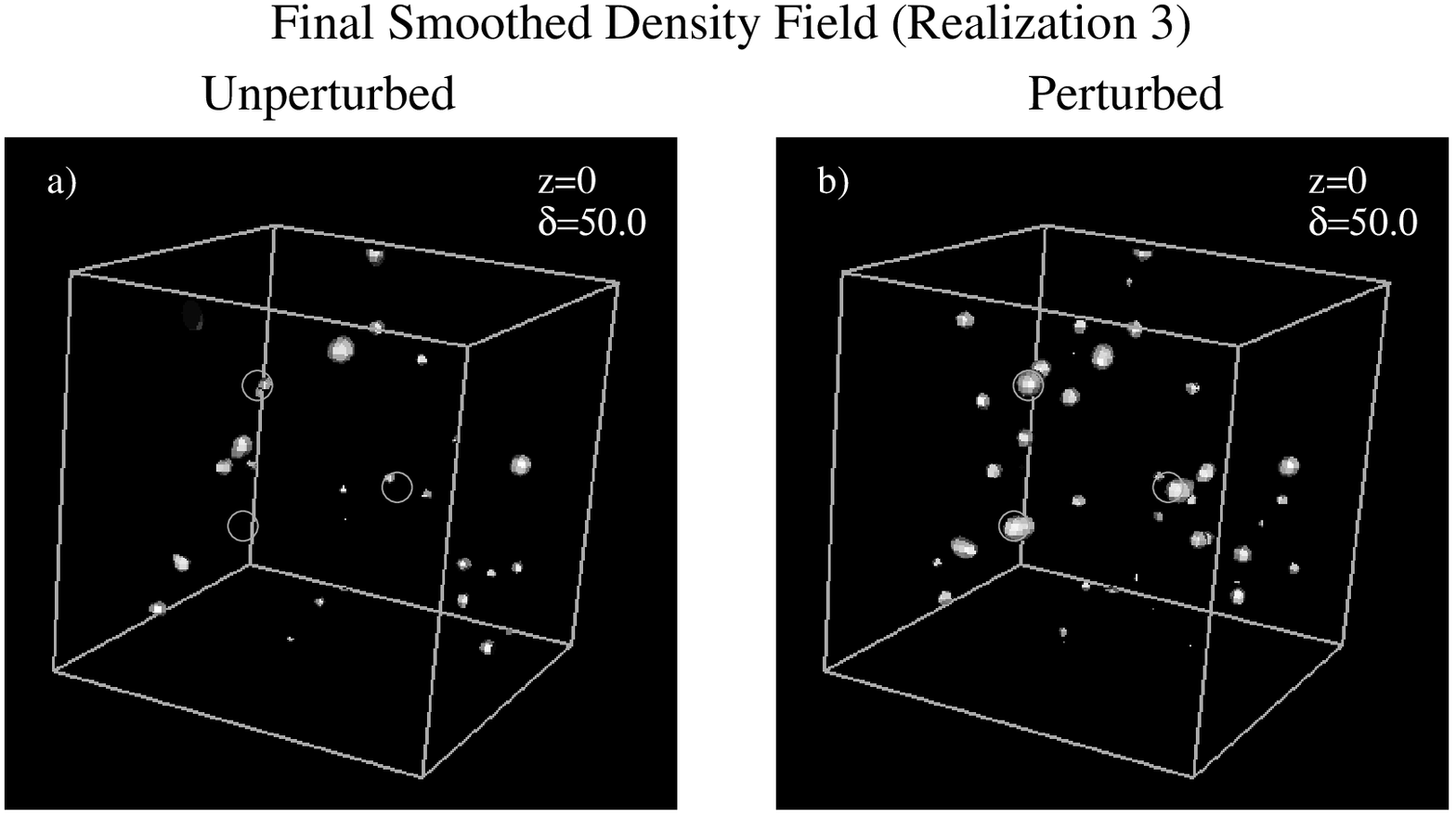,angle=-90,height=4in}}
\caption{Realization 2. As in Fig.~\ref{fg:sim1}, but for the third
realization.}
\label{fg:sim3}
\end{figure}

\begin{figure}
\centerline{\psfig{figure=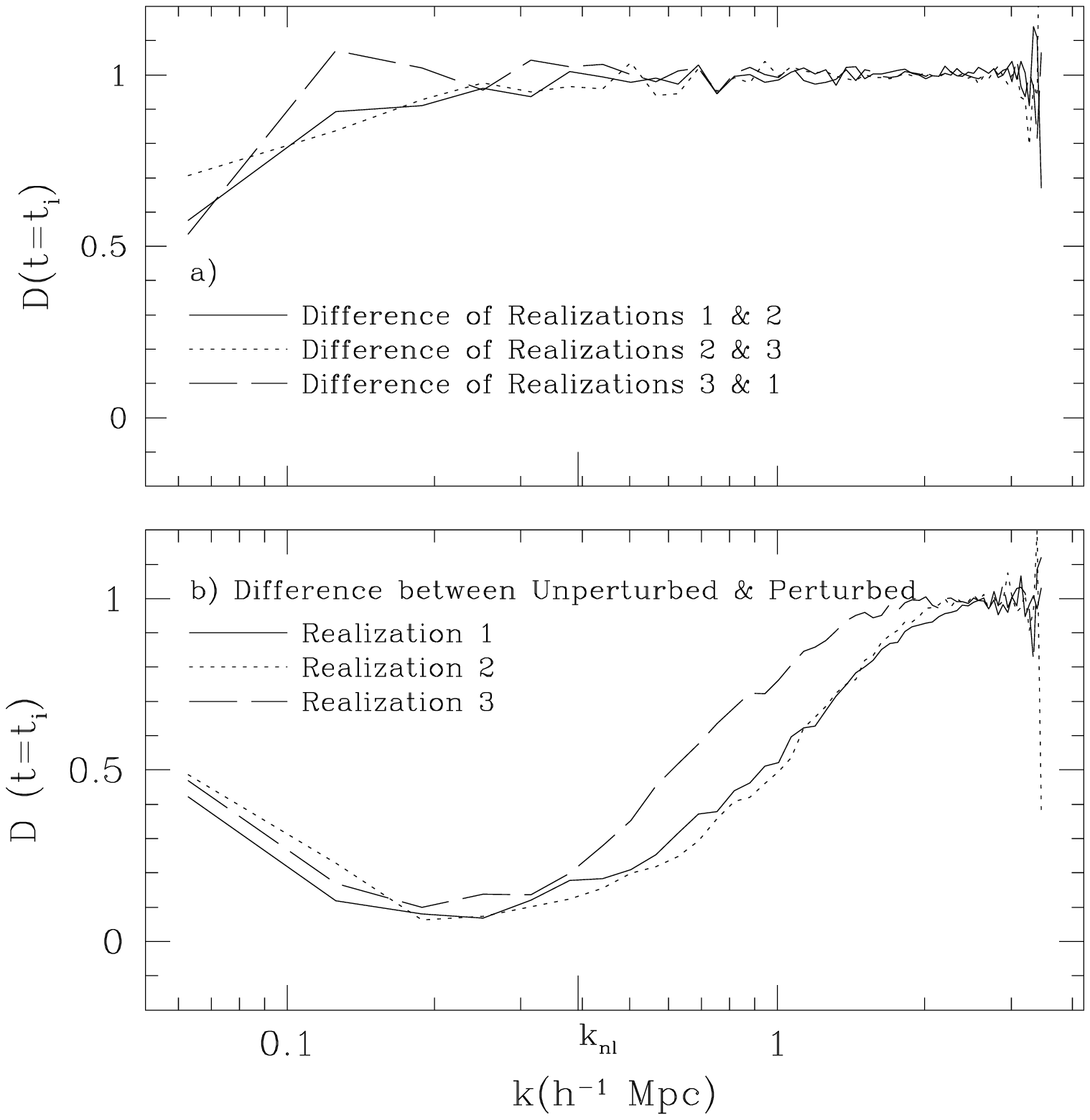,angle=0,height=6in}}
\caption{The Fourier Difference squared statistic for various
combinations of initial fields in test 1.  a) The three pairs of
perturbed density fields are completely uncorrelated with
one-another.  b)  Down to even mildly nonlinear scales, the perturbed
and unperturbed initial density fields have a very high correlation.}
\label{fg:dplot}
\end{figure}

Finally, we may consider a measure of how much a field needs to be
perturbed in order that it might satisfy the constraints.  In
Figure~\ref{fg:dplot}, we show the Fourier Difference statistic of a)
the three combinations of pairs of perturbed initial density fields
generated by PLA in this test, and b) the difference between the
perturbed and unperturbed initial conditions for the three
realizations.  Note that on all scales the three different
realizations are completely uncorrelated (up to cosmic variance).
However, on large and intermediate scales, the initial conditions
maintain much of their original structure. 

\subsubsection{A High-Resolution Realization}

\begin{figure}
\centerline{\psfig{figure=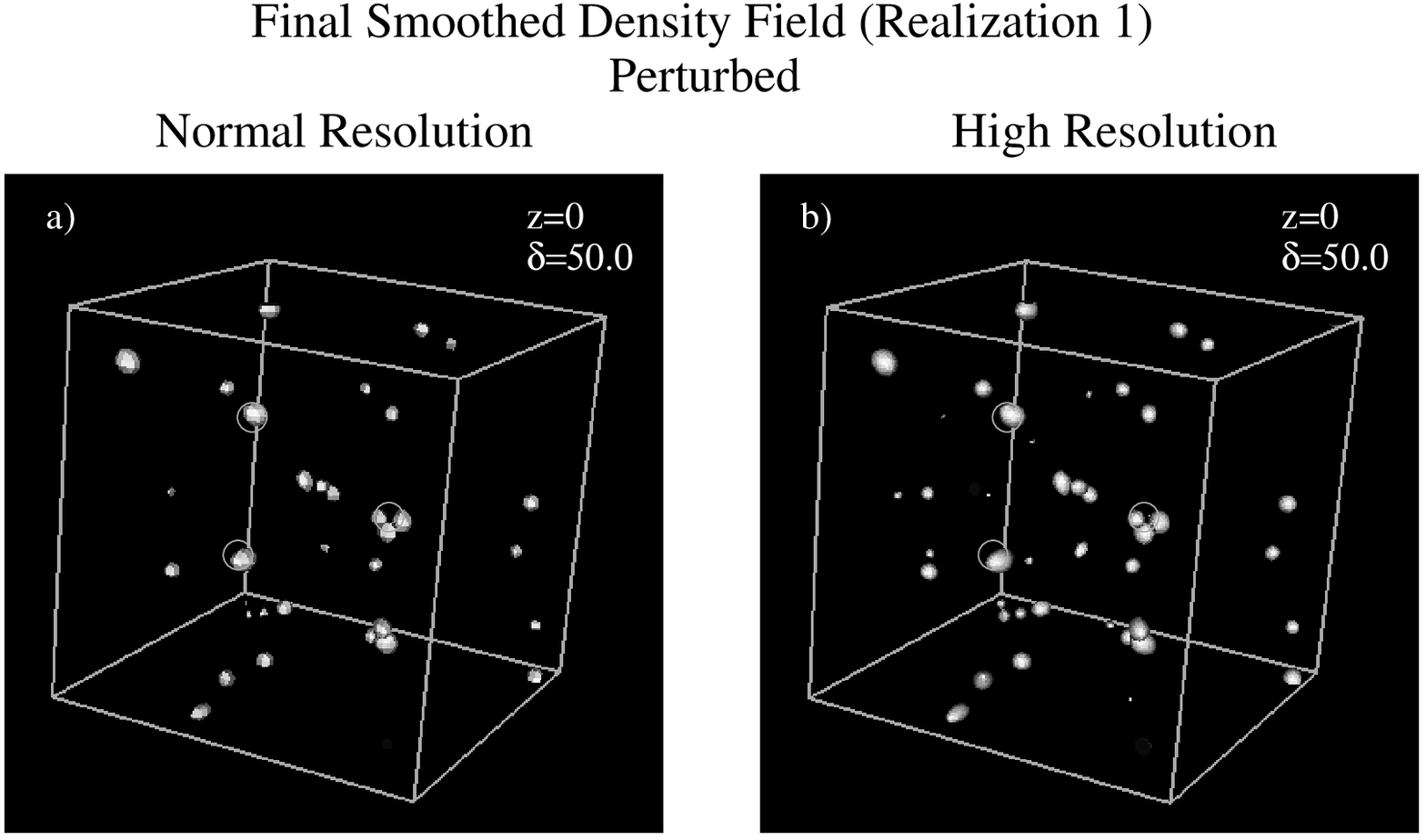,angle=-90,height=4in}}
\caption{A comparison of the the first realization run at a) normal
resolution, and b) high-resolution simulation, with small scale modes
filled in with known power spectrum.  The density contours and
smoothing scale are as in Fig.~\ref{fg:sim1}.}
\label{fg:highres}
\end{figure}

One of the great benefits of PLA is that once initial conditions have
been generated which satisfy a set of constraints, the field may be
Fourier decomposed, and large $k$ modes may be filled in using a known
primordial spectrum.  This new initial field will have higher
resolution than the original, and yet will reproduce all of the same
features on larger scales.  We have done this with the results of the
first realization in test
2, using $128^3$ particles and $256^3$ grid cells.  The results are shown
in Figure~\ref{fg:highres}.  

After running the high resolution initial conditions through the PM
code, the constraints are still satisfied to a tremendous degree, with
the three overdensities measuring $\overline{\delta}=$175, 195, \&
206, respectively.  In this respect, the consistency of the high- and
low-resolution simulations is quite telling.  Moreover, visual
inspection of the normal and high-resolution perturbed initial
conditions yield virtually identical results when smoothed on the same
scale.  In Figure~\ref{fg:highres}, we show a comparison of the
$\delta=50$ density contour of the two different resolutions, each
smoothed at $r=3h^{-1}$ Mpc.  The two fields appear almost identical,
suggesting that PLA is a viable technique for generating clusters in
specified positions, and then using those initial conditions as a seed
for a high-resolution study of the clusters.

\section{Future Goals}

This paper has largely concentrated on the method of using Perturbative
Least Action to generate initial conditions, and as a proof of concept,
we have generated initial conditions which beat PIZA in its ability
both to reproduce an initial density field and to match a final
density field. 

In addition to the basic method discussed here, future implementations
of the code will also incorporate redshift survey observations as well
as the potential for using an external, linearly evolving, tidal
field.  This will be extremely useful in studying semi-isolated
systems such as the Local Group of galaxies.  By modeling the Virgo
Cluster and the Great Attractor as perturbations on the local
potential field, we can realistically generate initial conditions and
model this system.  From there, we could ask meaningful questions
about infall history, dwarf galaxy statistics and so on.  Moreover, we
will have generated an initial density field which could be used as a
testbed for various N-body codes.  Finally, studies, such as those
done by Peebles (1989) based on the timing of the local group, could
be reproduced with extended halos in order to address the concerns
voiced by Branchini \& Carlberg (1995).

In the context of the Local Group, we will use PLA to explore
cosmological parameter space.  Basically, by running different
simulations with different cosmological parameters, and finding those
``Local Groups'' which best reproduce the statistical properties and
velocity field of the ``Local Group'', we will be able to 
independently estimate the true underlying cosmology.

In addition to highly nonlinear fields, we can use PLA to model
quasi-linear fields such as those observed in the IRAS survey.  An
initial power spectrum could then be generated which could be compared
to those produced using perturbation theory.  While groups have
investigated the evolution of the power spectrum using perturbation
theory (e.g. Jain \& Bertschinger 1994), PLA essentially evolves the
power spectrum to all orders, and moreover, preserves phase
information.  Using this approach, we will get a much stronger handle
on the primordial power spectrum on small scales.

Finally, in this paper we have described the case in which we have
observations constraining the final density distribution.  For a
redshift survey, however, one has a three-dimensional density field in
redshift space.  Giavalisco et al. (1993) point out that this can be
handled by performing a canonical transform on the basis functions.
Future implementations of the code will constrain the density field in
either real or redshift space.

\acknowledgements{We would like to thank P.J.E. Peebles, Michael
Strauss, Jerry Ostriker, Vijay Narayanan and Michael Vogeley for
helpful suggestions, and Michael Blanton for his invaluable
visualization software and useful comments on the manuscript.  DMG was
supported by an NSF Graduate Research Fellowship.  This research was
partially supported by NASA Theory Grant NAG \# 5-7154.}

\end{document}